\documentclass[superscriptaddress,preprint,showpacs,preprintnumbers,prb]{revtex4}
\usepackage{graphicx}
\usepackage{times}
\usepackage{epsfig}
\usepackage{color}
\usepackage{ulem}
\usepackage{dcolumn}
\usepackage{amsmath}
\begin{document}

\setlength{\textwidth}{17cm}

\title{Diffusion Quantum Monte Carlo study of the equation of state and
point defects in aluminum}

\author{Randolph Q. Hood}     \affiliation {Lawrence Livermore National 
Laboratory, Livermore, CA 94550, USA}
\author{P. R. C. Kent}        \affiliation {Center for Nanophase 
Materials Sciences and Computer Science and Mathematics Division, Oak Ridge
National Laboratory, Oak Ridge, TN 37831, USA}
\author{Fernando A. Reboredo} \affiliation {Materials Science and 
Technology Division, Oak Ridge National Laboratory, Oak Ridge, 
TN 37831, USA}
\begin{abstract}
  The many-body diffusion quantum Monte Carlo (DMC) method with
  twist-averaged boundary conditions is used to calculate the
  ground-state equation of state and the energetics of point defects
  in fcc aluminum, using supercells up to 1331 atoms. The DMC
  equilibrium lattice constant differs from experiment by 0.008 \AA
  \space or 0.2\%, while the cohesive energy using DMC with backflow
  wave functions with improved nodal surfaces differs by 27 meV. DMC
  calculated defect formation and migration energies agree with
  available experimental data, except for the nearest-neighbor
  divacancy, which is found to be energetically unstable in agreement
  with previous density functional theory (DFT) calculations. DMC and
  DFT calculations of vacancy defects are in reasonably close
  agreement. Self-interstitial formation energies have larger differences
  between DMC and DFT,  of up to 0.33eV, at the tetrahedral site. We also
  computed formation energies of helium interstitial defects where 
  energies differed by up to 0.34eV, also at the tetrahedral site. The
  close agreement with available experiment demonstrates that DMC can
  be used as a predictive method to obtain benchmark energetics of
  defects in metals.
\end{abstract}

\maketitle

\section{Introduction}
The mechanical properties of metals are dominated by the formation
and migration energies of defects. Experimentally it is often
difficult to measure desired defect properties directly, so it is
important to have accurate theoretical approaches to calculate defect
properties. The many-body diffusion quantum Monte Carlo (DMC) approach
\cite{foulkes01} is the most accurate method for systems with more than
$\approx 30$ electrons but has not previously been applied to defects
in metals. Until today, the most successful quantum mechanics-based
defect calculations in metals use density functional theory (DFT).
However, the approximate functionals used (i) lack sufficient specific
and universal accuracy, (ii) can not be systematically improved, and
(iii) there are now many approximate functionals to choose from, all
giving different results. Thus, DMC results are an ideal candidate to
wholly replace DFT when benchmark thermodynamic properties are required.
Furthermore, the relative energies of reference phases (bulk metals and
compounds) would be of great value in thermodynamic databases and in the
subsequent prediction of phase diagrams (e.g. CALPHAD and Thermocalc).

In semiconductors, DMC calculations of the formation energies of point
defects and surface energies have shown that important corrections to
DFT arise when electronic correlations are fully taken into account.
Deviations in formation energies of more than 1 eV were found in silicon
\cite{leung99} and diamond \cite{hood03}. Additionally, activation
energies of common chemical reactions obtained by DFT methods have been
shown to differ substantially from benchmark diffusion Monte Carlo (DMC)
values \cite{grossman97} and quantum chemical results. For metallic
systems, the size of the errors in DFT calculations of defects is largely
unknown, as more accurate benchmark calculations do not currently exist.
It is highly desirable to demonstrate the feasibility of DMC, with its
increased predictive accuracy, as a replacement of DFT for challenging
systems such as metals, particularly as computer power increases.

Aluminum is an ideal starting point for carrying out initial DMC calculations
of defects since it is one of the simplest metals with a close-packed
fcc structure that contains no 3d electrons. As a
result it is considered a prototype material for testing the validity of
theoretical calculations. Aluminum is well characterized experimentally
so there is an abundance of data available. There have been previous
quantum Monte Carlo calculations of the bulk properties of aluminum,
however, these were done using the less accurate variational Monte
Carlo \cite{gaudoin02a,gaudoin02b} and the calculations had large
statistical noise.

In this paper, we report well converged results for the bulk
properties of fcc aluminum using DMC. We explore a larger range
of volumes in order to compare the ground-state equation of state
calculated with DFT. Our DMC calculations of the defect properties
of aluminum include the simplest point defect, the vacancy for which
numerous DFT \cite{carling00,mattsson06,delczeg09} and experimental
results \cite{erhart91,hehenkamp94,fluss78,seeger71,desorbo59}
are available. We compute the nearest-neighbor divacancy binding
energy, a defect that DFT calculations \cite{carling00,uesugi03}
have found to be unstable. Since this instability is counter to both
experimental studies \cite{hehenkamp94,fluss84} and simple bond counting
arguments\cite{damask59} it is important to perform calculations of this
defect with DMC, a method that unlike DFT, does not rely on approximations
for exchange and correlation.

We also examine two other types of defects: First, self-interstitials
can arise due to irradiation with energetic particles, through plastic
deformation, or through their production in thermal equilibrium at
high temperatures. We have obtained DMC results for the formation
energies of the $\langle 100 \rangle$-dumbbell, the octahedral,
and the tetrahedral self-interstitials. Second, experimental
studies of irradiated aluminum show the presence of He bubbles
\cite{rajainmaki88,birtcher94,hamaguchi04}. The formation and growth of
helium bubbles can alter a material's mechanical properties through void
swelling, embrittlement, and surface blistering \cite{katoh03,vassen91}.
Since irradiation, through He implantation or transmutation, gives
rise to He atoms at substitutional or interstitial lattice sites, the
energetics of these types of defects are important. Presented here are
DMC calculated formation energies for He at the substitutional, octahedral
and the tetrahedral interstitial sites, which are compared with previous
DFT calculations \cite{yang08}. We demonstrate that twist-averaged
boundary conditions offer far superior statistics over DFT calculated
single-particle finite-size corrections for DMC calculations of defects,
and also remove a dependence on data from other methods. Overall the
calculations reported here used several million processor hours, which
is affordable on large computer clusters and supercomputers.

\section{Bulk Aluminum}

The DMC approach \cite{foulkes01} is a stochastic method for evolving
a wave function using the imaginary-time Schr\"{o}dinger equation.
In principle, and in contrast to most electronic structure methods,
the systematic errors can be measured and systematically reduced
\cite{umrigarprl2007,bajdichprl2010}. In practice, the most significant
sources of error are (i) the fixed-node and fixed-phase approximations,
a variational solution to the Fermion sign problem, (ii) adequately
sampling the Brillouin zone, which in contrast with insulators is a
significant problem in metals, and (iii) pseudopotential error and
corresponding locality error (although these may be avoided through
all-electron calculations \cite{esler10}).

\begin{figure}
\includegraphics[width=1.1\linewidth,clip=true]{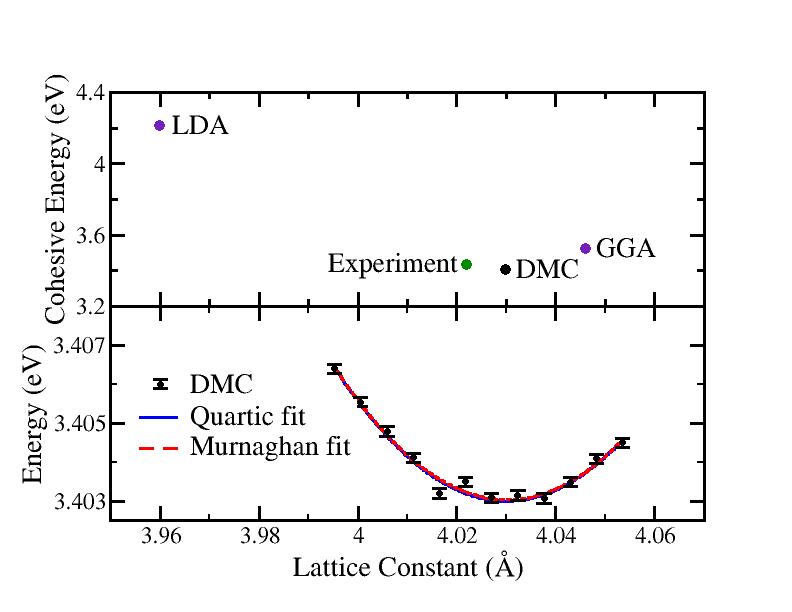}
\caption{(Color online) Upper: Comparison of equilibrium lattice constants
and cohesive energies for fcc aluminum calculated using DMC with backflow
(4.030(1) \AA, 3.403(1) eV), using DFT with LDA (3.960 \AA, 4.21 eV)
\cite{gaudoin02a} and GGA (PBE) (4.046 \AA, 3.52 eV), and experiment
(4.022 \AA, 3.43 eV) with zero-point energy and finite-temperature
effects removed \cite{gaudoin02a}. Lower: DMC energies for various
lattice constants, with a quartic and a Murnaghan \cite{murnaghan44}
fit. DMC data was obtained using twist-averaged boundary conditions
and extrapolated to infinite-sized supercells with single determinant
fixed-nodes. A shift of 0.063 eV was applied based on DMC calculations
using backflow wave functions (see text).}
\label{fg:energy_vs_a}
\end{figure}

For our DMC calculations we used the CASINO code \cite{needs10}, with
guiding wave functions formed by a product of Slater determinants
for up and down spin electrons and a Jastrow correlation function.
The single-particle orbitals in the determinants were obtained from
DFT calculations using the generalized gradient approximation (GGA)
for the exchange-correlation term since it should perform better than
the local density approximation (LDA) where the electron distribution
shows large spatial variations as it does at a vacancy in aluminum
\cite{carling00}. (For bulk aluminum GGA is more accurate than LDA, see
Fig.~\ref{fg:energy_vs_a}.) For GGA we used the Perdew-Burke-Ernzerhof
(PBE) form \cite{perdew96} rather than the Perdew-Wang-91 (PW91) form
\cite{perdew92} since it gives better values \cite{mattsson06} for the
vacancy formation energy. The DFT calculations were performed using the
plane-wave PWSCF code \cite{giannozzi09} with Troullier-Martins non-local
pseudopotentials. For the non-local pseudopotentials we used the locality
approximation \cite{mitas91} in the DMC calculations. The orbitals were
evaluated in DMC via real-space cubic splines \cite{williamson01}. For
the DMC simulations we used 5280 walkers with a time step of 0.01 au,
which our test calculations revealed gave time-step errors in Table
\ref{energies} of less than 0.01 eV.

Calculations of solids using DMC require finite simulation supercells
with periodic boundary conditions imposed on the Hamiltonian. This leads
to finite-size effects that can be divided between single-particle and
many-body contributions. In a metal the DMC kinetic energy contains
large single-particle contributions due to the sharp Fermi surface,
which impacts our calculations in two ways.

First, in a real metal the number of orbitals with energies below the
Fermi level is usually not equal to the number of electrons required in
the simulation cell. In DFT calculations one can use partial occupations
of these orbitals to create a closed shell configuration guaranteeing
that the charge density has the correct symmetry. In a DMC calculation
with a guiding wave function containing a single determinant for the
spin up and for the spin down electrons the use of partial occupations
is not an option. Gaudoin, et. al. \cite{gaudoin02b} found differences
in aluminum as large as 0.1 eV/atom in variational Monte Carlo total
energies depending on the occupations of the orbitals at the Fermi level.
As our calculations show in Fig.~\ref{fg:energy_vs_a} for aluminum 
an error on the order of 0.1 eV/atom is too large to accurately discern
the equilibrium lattice constant.

\begin{figure}
\includegraphics[width=1.0\linewidth,clip=true]{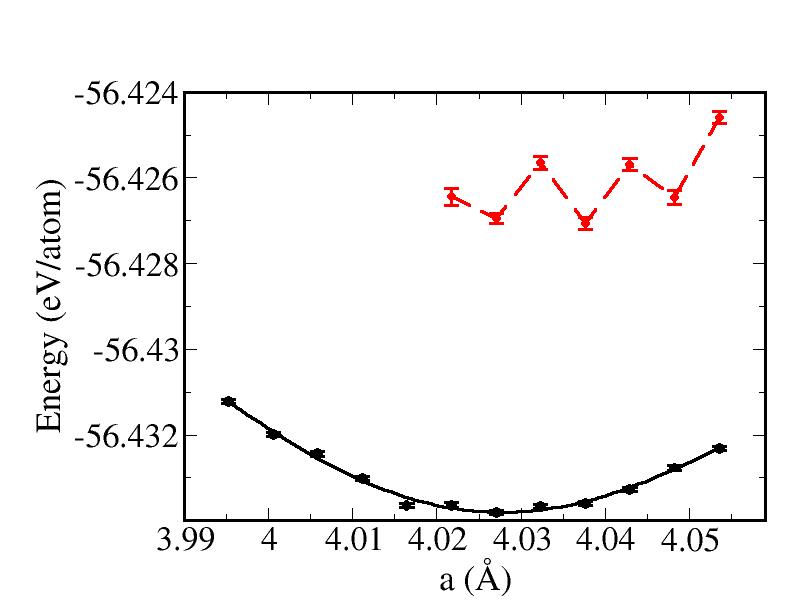}
\caption{(Color online) The black circles with error bars are the
calculated DMC ground-state energies of bulk fcc aluminum obtained using
twist-averaged boundary conditions with $\text{10x10x10}$ twists. The red
diamonds with error bars are the calculated DMC ground-state energies
of bulk fcc aluminum obtained without using twist-averaged boundary
conditions. The DFT corrections $E_{\infty}^{DMC} = E_{N}^{DMC} +
E_{\infty}^{DFT} - E_N^{DFT}$ were added to the results that did not
use twist-averaged boundary conditions. Both sets of calculations were
done using $\text{5x5x5}$ supercells containing $N=125$ atoms.}
\label{fg:bulk_w_wo_twist}
\end{figure}

A second complication for DMC calculations of metals arises as the system
size is varied, which can produce band crossings as energy levels pass
through the Fermi level. This can create discontinuous changes in the
nodal surface of the guiding wave function as the symmetry of the occupied
orbitals change, leading to discontinuities in the DMC energy as shown in
Fig.~\ref{fg:bulk_w_wo_twist}. For simulation cells containing 64 or 125
atoms we found that single-particle DFT corrections $E_{\infty}^{DMC} =
E_{N}^{DMC} + E_{\infty}^{DFT} - E_N^{DFT}$ were ineffective in removing
the large discontinuities we found in calculated DMC energies as the
volume was varied. However, the use of well-converged twist-averaged
boundary conditions \cite{lin01,teweldeberhan10} was effective in
producing smooth energy versus volume curves at these systems sizes as
shown in Fig.~\ref{fg:bulk_w_wo_twist}. Our calculations with 64 and
125 atoms used $\Gamma$-point centered grids with $\text{13x13x13}$
and $\text{10x10x10}$ twists, respectively.

\begin{figure}
\includegraphics[width=1.0\linewidth,clip=true]{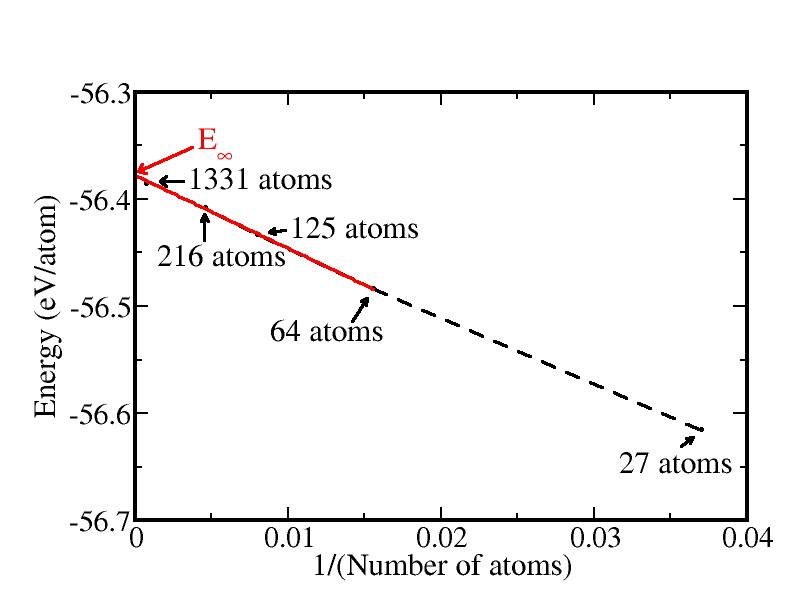}
\caption{(Color online) The circles are calculated DMC ground-state
energies of fcc aluminum with lattice constant a=4.046 \AA \space using
twist-averaged boundary conditions for simulation cells containing 27,
64, 125, 216 atoms, and 1331 atoms, with a $\Gamma$-point centered grid of
$\text{17x17x17}$, $\text{13x13x13}$, $\text{10x10x10}$, $\text{9x9x9}$,
and $\text{1x1x1}$ twists, respectively. The statistical error bars are
smaller than the circles. The dashed line is a guide to the eye. The solid
red line is a fit to the data using Eq.~(\ref{eq:finite_size}) where N,
the number of atoms, is 64, 125, and 216. This fit has a correlation
coefficient of -0.9997.}
\label{fg:finite_size_effects}
\end{figure}

With twist-averaged boundary conditions the remaining finite-size effects arise
from many-body contributions \cite{drummond08} that scale as
\begin{equation}
E_{\infty} = E_N + c/N .
\label{eq:finite_size}
\end{equation}
As shown in Fig.~\ref{fg:finite_size_effects} this equation agrees
well with our DMC energies using twist-averaged boundary conditions for
simulation cells containing between 27 and 1331 atoms.

We used Eq.~(\ref{eq:finite_size}) to obtain infinite-size extrapolated
DMC total energies from calculations using 64 and 125 atom simulation
cells with twist-averaged boundary conditions using a range of lattice
constants shown in Fig.~\ref{fg:energy_vs_a}. The energy points were fit
to a quartic and a Murnaghan equation of state \cite{murnaghan44}. Both
fits yielded an equilibrium lattice constant of 4.030(1) \AA. This
compares well with the experimental value, 4.022 \AA, with zero-point
energy and finite-temperature effects removed \cite{gaudoin02a}. In
contrast DFT calculations with the local density approximation (LDA)
and GGA (PBE) yield 3.960 \AA \space and 4.046 \AA, respectively.

For the DMC cohesive energy we initially obtained 3.341(1) eV using the
fixed-nodes defined by the GGA orbitals, compared with the experimental
value of 3.43 eV with zero-point energy and finite-temperature effects
removed, 4.21 eV using LDA, and 3.52 eV using GGA (PBE). Although the
fixed-node DMC results with GGA nodes are already the most accurate,
to assess the possible nodal error we also performed DMC calculations
with twist-averaged boundary conditions and extrapolation to an
infinite-sized supercell using optimized backflow wave functions
(with backflow transformations that contained electron-electron,
electron-nuclei, and electron-electron-nuclei terms \cite{rios06}) at
the optimum lattice constant and also for an isolated atom. Backflow
wave functions can be substantially more accurate than single determinant
non-backflow wave functions, typically yielding an additional few percent
DMC correlation energy in atomic calculations and nearly 100 percent
of the correlation energy in the homogeneous electron gas\cite{rios06}.
However, since backflow is too expensive to apply routinely, we have
used the backflow result, similar to how corrections for all-electrons
have been applied \cite{esler10}, to shift our single determinant
fixed-node energies in Fig.~\ref{fg:energy_vs_a}. The backflow cohesive
energy is 3.403(1) eV. A complete backflow evaluation of the lattice
constant might further reduce the residual differences from experiment.

We expect the backflow correction in metallic systems with atomic number
higher than aluminum to be at least as large as our computed correction
in aluminum. This indicates that to obtain cohesive energies to better
than 0.1eV -- other than by fortuitous error cancellation -- backflow or
other nodal optimization must be considered.

\begin{figure}
\includegraphics[width=1.1\linewidth,clip=true]{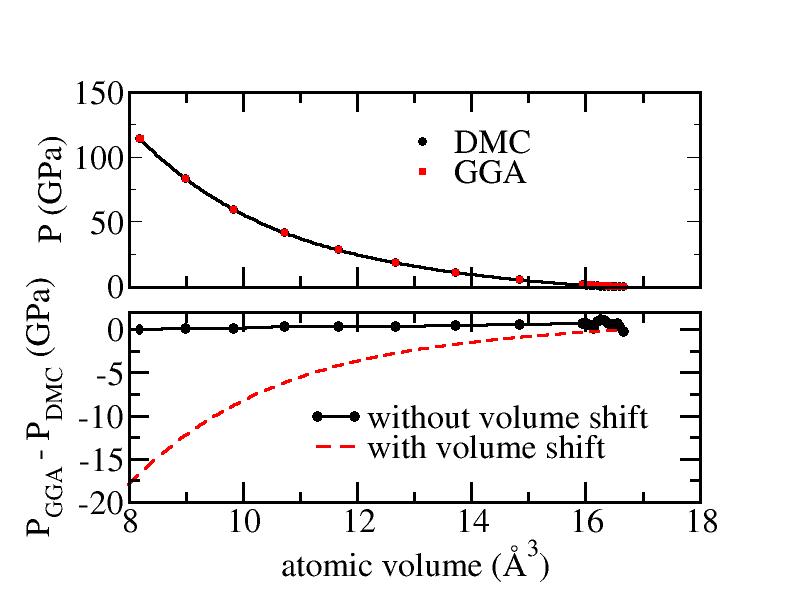}
\caption{(Color online) Upper: The red squares show the calculated
pressures of fcc aluminum using DFT with GGA (PBE). The black circles are
pressures obtained from a Murnighan fit of total energy DMC calculations
performed at a range of atomic volumes. At this scale the GGA and DMC
data are indistinguishable. Lower: The black solid line is the difference
in pressure between the GGA and the DMC calculations at the same range
of atomic volumes. The dashed red line shows the difference in pressures
between the GGA and the DMC calculations after the GGA energy-volume
curve as been shifted so that the equilibrium volumes of the GGA and
DMC calculations coincide.}
\label{fg:pressure_plots}
\end{figure}

We performed additional DMC total energy calculations of bulk aluminum
for atomic volumes smaller than those shown in Fig.~\ref{fg:energy_vs_a}
by following the same procedure of using Eq. (1) to obtain infinite-size
extrapolated DMC total energies from calculations using 64 and 125 atom
simulation cells with converged twist-averaged boundary conditions. A
Murnaghan fit of the energy-volume DMC data was used to obtain the
pressure for a range of atomic volumes. These calculated DMC pressures
are shown in the upper part of Fig.~\ref{fg:pressure_plots} along
with pressures calculated with DFT using GGA (PBE). At this scale the
differences between the DMC and GGA pressures are not visible. The solid
black curve in the lower part of Fig.~\ref{fg:pressure_plots} shows
the difference in pressures between GGA and DMC. A common procedure for
constructing an equation of state of a material at low temperatures is to
shift the computed equilibrium lattice constant so that it coincides
with the experimental equilibrium lattice constant\cite{correa08}. 
Following a similar procedure of shifting the GGA calculated energy-volume
curve so that the GGA equilibrium lattice constant of 4.046 \AA\;
coincides with the DMC equilibrium lattice constant of 4.030(1) \AA\; yields 
a pressure difference between the GGA and the DMC equation of states that
increases markedly at smaller atomic volumes as shown in the red dashed
line in Fig.\ref{fg:pressure_plots}. This demonstrates that the applying
a rigid shift to DFT equation of state calculations can result in larger
errors than if a shift is not applied.

\section{Defects in aluminum}

The results of our calculations of point defects are presented in Table
\ref{energies}. The atomic positions were taken from complete structure
and volume relaxed DFT calculations using GGA at zero pressure. The
calculated GGA and DMC vacancy formation and migration energy agree
with experiment. However, GGA no longer agrees with experiment when
``surface'' corrections \cite{carling00,sandberg02,mantina08} of
0.15 eV and 0.05 eV are added to the GGA (PBE) functional producing
0.82 eV and 0.65 eV for the vacancy formation and migration energy,
respectively. Previous GGA (PBE) calculations \cite{mattsson06} of the
vacancy formation energy without the ``surface'' correction using 4x4x4
supercells yielded values of 0.61 eV and 0.63 eV with norm-conserving
and projector augmented-wave pseudopotentials, respectively. This
difference with our GGA results is likely partially due to finite-size
effects since we obtained 0.64 eV for the vacancy formation energy using
a $\text{4x4x4}$ supercell. The GGA results in Table \ref{energies}
correspond to calculations using finite-size converged $\text{7x7x7}$
supercells. Convergence of the GGA defect structures was established
by computing the energies using $\mathrm{4x4x4}$, $\mathrm{5x5x5}$,
$\mathrm{6x6x6}$, and $\mathrm{7x7x7}$ supercells. The DMC results
in Table \ref{energies} were done using $\mathrm{5x5x5}$ supercells
with $\Gamma$-point centered grids of $\mathrm{10x10x10}$ twists. The
finite-size errors for our DMC calculations are likely to be small since
the largest GGA energy difference among all of the defects comparing
$\mathrm{5x5x5}$ and $\mathrm{7x7x7}$ supercells was 0.02 eV. Shown in
Table \ref{finite_size_energies} are GGA and DMC data demonstrating the
convergence with supercell size for all the defects considered.

\begin{table}
\caption{Energies (enthalpies at zero pressure) for point defects in
fcc aluminum (eV) calculated using GGA (PBE) and DMC. Formation and
migration energies for a single vacancy are denoted $H_v^f$ and $H_v^m$,
respectively. Shown in braces are the GGA values with ``surface''
corrections \cite{carling00,sandberg02} added. Formation and binding
energies for a nearest-neighbor divacancy are denoted $H_{2v}^f$ and
$H_{2v}^b$, respectively. $H_d^f$, $H_o^f$ and $H_t^f$ are the formation
energies for the $\langle 100 \rangle$-dumbbell, the octahedral, and
the tetrahedral self-interstitials, respectively. Formation energies
for a He impurity at a substitutional, octahedral, and tetrahedral
sites are denoted $H_{He(s)}^f$, $H_{He(o)}^f$, and $H_{He(t)}^f$,
respectively. Statistical errors bars for all DMC energies are 0.01
eV. GGA and DMC calculations were done using $\text{7x7x7}$ and
$\text{5x5x5}$ supercells, respectively, with atomic positions taken
from complete structure and volume relaxed GGA calculations.}
\label{energies}
\begin{ruledtabular}
\begin{tabular}{lcdc}
               & GGA & \multicolumn{1}{r}{DMC}   & Experiment \\ \hline
$H_{v}^f$ & 0.67 \{0.82\} & 0.67 & 0.67(3) \cite{erhart91},
0.67 \cite{hehenkamp94}, 0.66(2) \cite{fluss78} \\
$H_{v}^m$ & 0.60 \{0.65\} & 0.60 & 0.62 \cite{seeger71},
0.61(3) \cite{erhart91}, 0.65(6) \cite{desorbo59} \\
$H_{2v}^f$ & 1.37 & 1.44 & 1.17(7)\footnote{Computed using $H_{v}^f$ from Ref.
[\onlinecite{erhart91}] and $H_{2v}^b$ from Ref. [\onlinecite{doyama64}]} \\
$H_{2v}^b$ & -0.03 & -0.10 & 0.17(5) \cite{doyama64},
0.20 \cite{hehenkamp94} \\
$H_{d}^f$ & 2.70 & 2.94 & 3.0 \cite{erhart91}, 3.2(5) \cite{schilling78}\\
$H_{o}^f$ & 2.91 & 3.13 & \\
$H_{t}^f$ & 3.23 & 3.56 &  \\
$H_{He(s)}^f$ & 1.63 & 1.72 & \\
$H_{He(o)}^f$ & 3.26 & 3.58 & \\
$H_{He(t)}^f$ & 3.33 & 3.67 & \\
\end{tabular}
\end{ruledtabular}
\end{table}

\begin{table}
\caption{Energies (enthalpies at zero pressure) for point defects in fcc
aluminum (eV) calculated using density functional theory (DFT) with GGA
(PBE) functional and diffusion Monte Carlo (DMC). The symbols for the
formation and migration energies are the same as those used in Table
\ref{energies}. Except where indicated the statistical errors bars for
the DMC energies are 0.01 eV. The GGA and DMC calculations were done
with atomic positions taken from complete structure and volume relaxed
GGA calculations. For the GGA calculations we used $\text{4x4x4}$,
$\text{5x5x5}$, $\text{6x6x6}$, and $\text{7x7x7}$ supercells
with $\Gamma$-centered grids of $\text{13x13x13}$, $\text{10x10x10}$, 
$\text{9x9x9}$, and $\text{8x8x8}$ kpoints, respectively. For the DMC
calculations we used $\text{4x4x4}$ and $\text{5x5x5}$ supercells. For
DMC without twist averaging we applied the single-particle DFT corrections
$E_{\infty}^{DMC} = E_{N}^{DMC} + E_{\infty}^{DFT} - E_N^{DFT}$. For the
DMC calculations with twist averaging we did not apply DFT corrections
and used $\Gamma$-centered grids with $\text{13x13x13}$ twists and
$\text{10x10x10}$ twists for the $\text{4x4x4}$ and $\text{5x5x5}$
supercells, respectively.  Entries in the table that were not determined
are denoted N.D.}
\label{finite_size_energies}
\begin{ruledtabular}
\begin{tabular}{lddddd|dd|dd}
              & \multicolumn{4}{c} {GGA} & & \multicolumn{2}{c|} {DMC with DFT corrections} & \multicolumn{2}{c} {DMC with twist averaging} \\ \hline
supercell              & \multicolumn{1}{c}{$\;\;\;\text{4x4x4}$} & \multicolumn{1}{c}{$\;\;\;\text{5x5x5}$} & \multicolumn{1}{c}{$\;\;\;\text{6x6x6}$} & \multicolumn{1}{c}{$\;\;\;\text{7x7x7}$} & &\multicolumn{1}{c}{$\;\;\;\text{4x4x4}$} & \multicolumn{1}{c|}{$\;\;\;\text{5x5x5}$} &\multicolumn{1}{c}{$\;\;\;\text{4x4x4}$} & \multicolumn{1}{c}{$\;\;\;\text{5x5x5}$} \\ \hline
$H_{v}^f$       & 0.64 & 0.65 & 0.67 & 0.67 & & 0.58    & 0.81    & 0.66 & 0.67 \\
$H_{v}^m$       & 0.59 & 0.59 & 0.60 & 0.60 & & 0.50    & 0.40(3) & 0.54 & 0.60 \\
$H_{2v}^f$      & 1.36 & 1.35 & 1.36 & 1.37 & & 1.31    & 1.29(3) & 1.47 & 1.44 \\
$H_{2v}^b$      &-0.05 & -0.05& -0.02& -0.03& &-0.15(2) & 0.33(5) &-0.15 &-0.10\\
$H_{d}^f$       & 2.82 & 2.72 & 2.72 & 2.70 & & N.D.    & N.D.    & N.D. & 2.94 \\
$H_{o}^f$       & 2.93 & 2.90 & 2.94 & 2.91 & & 2.90    & 2.75    & 3.18 & 3.13 \\
$H_{t}^f$       & 3.53 & 3.25 & 3.27 & 3.23 & & 3.27    & 3.27    & 3.60 & 3.56 \\
$H_{He(s)}^f$   & 1.61 & 1.61 & 1.62 & 1.63 & & 1.41    & 1.98    & 1.64 & 1.72 \\
$H_{He(o)}^f$   & 3.26 & 3.24 & 3.27 & 3.26 & & 3.29    & 3.45    & 3.53 & 3.58 \\
$H_{He(t)}^f$   & 3.44 & 3.35 & 3.35 & 3.33 & & 2.89    & 3.55    & 3.71 & 3.67 \\
\end{tabular}
\end{ruledtabular}
\end{table}

For the nearest-neighbor divacancy with DMC we obtain a negative
binding energy, -0.10 eV, which implies that two isolated vacancies are
energetically preferred to a nearest-neighbor divacancy. This agrees
with previous DFT calculations which also found a similar negative
binding energy, -0.05 eV using \cite{uesugi03} LDA and -0.08 eV using
\cite{carling00} GGA (PW91). Thus the disagreement between previous
calculations and the original interpretation of experimental data
\cite{hehenkamp94,fluss84}, which gave positive binding, is likely not
a result of DFT approximations. Our results are consistent with the
reinterpretation \cite{carling00} of the data.

Experimentally \cite{schilling78} the $\langle 100 \rangle$-dumbbell
was found to be the lowest energy self-interstitial in aluminum. Of
the self-interstitials investigated we found that the $\langle 100
\rangle$-dumbbell has the lowest formation energy. The calculated
formation energy was 2.94 eV using DMC and 2.70 eV using GGA. Our DMC
value agrees with the experimental estimates of 3.0 \cite{erhart91} and
3.2(5) eV \cite{schilling78}. For the $\langle 100 \rangle$-dumbbell we
obtain a relaxation volume of 2.25, which agrees closely with experimental
estimates of 1.9(4) and 1.7(4) \cite{schilling78}. Our GGA (PBE) result
is larger than a previous GGA (PW91) result \cite{sandberg02} of 2.43
eV. For the self-interstitials we see differences between the calculated
DMC and GGA formation energies as large as 0.24 eV.

The DMC formation energy for a He substitutional defect is 1.72 eV, while
the energies for He interstitials are larger than 3 eV. The ordering
of these energies is consistent with sites with larger free volumes
having lower energies. Previous DFT calculations \cite{yang08} using GGA
(PW91) obtained 1.53, 3.18, and 3.20 eV for He at the substitutional,
octahedral, and tetrahedral site, respectively. Comparing our GGA and
DMC calculations we see differences between 0.09 and 0.34 eV for the He
impurity. Similar to the self-interstitials, the GGA exchange-correlation
errors are larger for these defects than for the vacancy.

Relying on DFT corrections to minimize the single-particle finite-size
errors instead of twist averaging, yielded poorer convergence for defect
energies. For example, the difference in vacancy formation energies
between $\text{4x4x4}$ and $\text{5x5x5}$ supercells was 0.23 eV compared
with 0.01 eV with twist averaging, while the result for divacancy
binding was reversed as shown in Table \ref{finite_size_energies}.
DMC calculations in larger $\text{6x6x6}$ supercells would be an order of
magnitude more expensive, and may still be inferior to the twist-averaged
results. The use of twist averaging is essential in metals for defects
and excitations in which the fractional change in the total energy due
to the presence of the defect or excitation is inversely proportional
to the number of atoms in the supercell, i.e., ``1/N'' effects.

\section{Conclusions}

In summary, DMC with twist-averaged boundary conditions can be used to
obtain an accurate equation of state of aluminum. Our DMC results confirm
previous DFT calculations that the nearest-neighbor divacancy is unstable in
aluminum. Our calculated formation and migration energies of point defects
show excellent agreement with available experiment, demonstrating that DMC
can be used to obtain benchmark energetics of defects in metals and used
as a baseline where no experiment is available.

\begin{acknowledgments}
We thank J. Dubois and J. Kim for useful discussions.  This material is
based upon work supported as part of the CDP, an Energy Frontier Research
Center funded by the U.S. Department of Energy (DOE), Office of Science,
Office of Basic Energy Sciences under Award Number ERKCS99. This work
performed under the auspices of the U.S. DOE by LLNL under Contract
DE-AC52-07NA27344.  Computing support for this work came from the LLNL
Institutional Computing Grand Challenge program. Research performed at
the Materials Science and Technology Division and the Center of Nanophase
Material Sciences at ORNL was sponsored by the Division of Materials
Sciences and the Division of Scientific User Facilities U.S. DOE.
\end{acknowledgments}

\end{document}